\def\e{{\rm e}}
\def\del{\partial}
\def\half{{1\over2}}

\def\abs#1{{\left|{#1}\right|}}

\def\ket#1{|{#1}\rangle}
\def\bra#1{\langle {#1}|}
\def\del{\partial}
\def\dslash{\del\kern-0.55em\raise 0.14ex\hbox{/}}
\def\hbar{h\kern-0.50em\raise 0.75ex\hbox{-}}

\newcommand{\norm}[2]{\langle{#1}|{#2}\rangle}
\newcommand{\PRD}[3]{Phys. Rev. {\bf D{#1}} (19{#2}) {#3}} 
\newcommand{\aPRD}[3]{Phys. Rev. {\bf D{#1}} (20{#2}) {#3}} 
 
\newcommand{\NPB}[3]{Nucl. Phys. {\bf B{#1}} (19{#2}) {#3}} 
\newcommand{\PLB}[3]{Phys. Lett. {\bf B{#1}} (19{#2}) {#3}} 
\newcommand{\PTP}[3]{Prog. Theor. Phys. {\bf {#1}} (19{#2}) {#3}} 
\newcommand{\ANN}[3]{Ann. Phys. {\bf {#1}} (19{#2}) {#3}} 
\newcommand{\PR}[3]{Phys. Rep. {\bf {#1}} (19{#2}) {#3}}

\newcommand{\JMP}[3]{J. Math. Phys. {\bf {#1}} (19{#2}) {#3}}
\newcommand{\ZP}[3]{Zeit. Phys. {\bf C{#1}}(19{#2}) {#3}} 
\newcommand{\MPL}[3]{Mod. Phys. Lett. {\bf A{#1}}(19{#2}) {#3}}
\newcommand{\hg}{\hat G}
\newcommand{\hw}{\hat W}
\newcommand{\hwd}{{\hat W}^{\dagger}}

\newcommand{\hq}{\hat Q}
\newcommand{\hqb}{\hat{\bar Q}}

\newcommand{\hxi}{\hat\xi}
\newcommand{\hxib}{\hat{\bar\xi}}
\documentclass[12pt]{article}
\begin{document}
\title{Quantization Ambiguity \\and \\
Supersymmetric Ground State Wave Functions}
\author{Kazunori Takenaga \vspace{1cm}$^{}$\thanks {email: 
takenaga@alf.nbi.dk}\\ 
\it {Niels Bohr Institute, Copenhagen University,}\\ 
{\it Blegdamsvej 17, DK-2100 Copenhagen $\emptyset$ Denmark}}
\date{} 
\maketitle
\baselineskip=18pt
\vskip 3cm
\begin{abstract}
Supersymmetric ground state wave functions of 
a model of supersymmetric quantum mechanics on $S^1$ (supersymmetric 
simple pendulum) are studied. Supersymmetry can be broken due to the 
existence of an undetermined parameter, which is interpreted 
as a gauge field and appears as 
a firm consequence of quantization on a space with 
a nontrivial topology such as $S^1$. The breaking does not depend on 
the leading term of the superpotential, contrary to the usual case.
The mechanism of supersymmetry breaking is similar to that through  
boundary conditions of fields in supersymmetric
quantum field theory on compactified space. 
The supersymmetric harmonic oscillator is 
realized in the limit of the infinite 
radius of $S^1$ with the strength of the oscillator being constant.
\end{abstract}
\vskip 2cm
\begin{flushleft}
NBI-HE-00-16\\
March 2000\\
\end{flushleft}
\addtolength{\parindent}{2pt}
\newpage
\section{Introduction}
In this paper we study how the quantization ambiguity, which implies 
that quantization on space with a nontrivial topology such 
as $S^1$ inevitably yields an undetermined parameter 
into the theory \cite{tsutsui}, affects the supersymmetric ground state wave 
functions 
of a model of supersymmetric quantum mechanics on $S^1$. 
\par
Quantum mechanics on $S^1$ was studied for the first time in the path-integral 
formalism \cite{schul1}\cite{laid}. We observe that the aforementioned 
parameter appears as a phase factor in the Feynman kernel due to the 
nontrivial topology of configuration space and, as a result, there 
are many distinct propagators labeled by the parameter. 
One can also consider the same effect in 
the Lagrangian by adding a total derivative term whose coefficient is given 
by the parameter. Accordingly, the canonical momentum is shifted by the 
amount of the parameter, so that it can be interpreted as a constant 
gauge field. The total derivative term has physical implications 
at the quantum level for space with nontrivial topology \cite{schul2}.
\par
Quantization on $S^1$ is much different from that on one-dimensional
Euclidean space. In the language of canonical formalism, the latter 
case is that the representation of the canonical algebra is uniquely 
determined up to a unitary equivalent representation. There is essentially one 
quantum mechanics on the space. In the former case, however, there is
an infinite number of inequivalent representations of the 
fundamental algebra, which is introduced as a generalization of the canonical
algebra by Ohnuki and Kitakado \cite{ok} in order to formulate 
quantum mechanics on $S^1$.
As a result, there exists various quantum mechanics on $S^1$.  
\par
The various quantum mechanics on $S^1$ are parametrized 
by the undetermined parameter. The parameter is interpreted as a 
constant gauge field \cite{ok}\cite{tanimura}. 
The gauge field can exist and has effects on observables at the 
quantum level. It is the existence of the gauge field that leads to the  
various quantum mechanics on $S^1$. 
It may be interesting to study the possible 
effects of the gauge field on the supersymmetric 
ground state (zero-energy state) wave 
functions of supersymmetric quantum mechanics on $S^1$. 
The zero-energy state wave functions can be obtained in closed form 
\cite{claud} because, thanks to the 
supersymmetry algebra satisfied by the system, the wave functions are  
obtained by solving simple first-order equations in many cases. 
Therefore, it may be possible to study 
the effects as analytically as possible.   
\par
Supersymmetric quantum mechanics has been studied in great 
detail and applied to many physics fields \cite{cooper}. Actually, it 
provides us with an example of the dynamical supersymmetry 
breaking by instantons
in certain models \cite{witten}\cite{holton}\cite{freedman}\cite{khare}.  
In those models, the normalizability of the supersymmetric ground state 
wave function crucially depends on the leading term in the 
superpotential, by which we determine whether or not 
the supersymmetry is broken. The semiclassical 
instanton approximation has been used to estimate
the ground state energy for the system with broken 
supersymmetry\footnote{Estimating the ground state energy is a subtle problem.
Actually, it is reported in \cite{abbott1}\cite{abbott2} that the instanton 
calculation is very limited though it gives us an excellent estimation 
in some cases and the 
breakdown of symmetry is caused by the interplay of perturbative 
and nonperturbative effects.}.
\par
In this paper we will find an another mechanism 
of supersymmetry breaking. The very existence of the gauge field twists
boundary conditions of supersymmetric ground state wave functions.
For certain values of the gauge field, the wave functions do not satisfy
a required periodic boundary condition and become
unphysical though they are normalizable. 
The supersymmetry breaking does not depend on the 
structure of the superpotential, unlike the usual
supersymmetry breaking discussed in supersymmetric quantum mechanics. 
Supersymmetry can be broken by the
gauge field, that is, the quantization ambiguity. 
\par
In the next section we shall introduce a model of supersymmetric 
quantum mechanics on $S^1$ after reviewing briefly the 
quantum mechanics on $S^1$
formulated by Ohnuki and Kitakado.  And then, we shall 
discuss how the gauge field affects the supersymmetric ground state wave 
functions and how it yields supersymmetry breaking. 
We shall also study an infinite limit of the radius of $S^1$. 
The supersymmetric
harmonic oscillator is realized in the limit with the 
strength of the oscillator being constant. 
The final section is devoted to conclusions and a discussion, where 
we shall also discuss the similarities between our mechanism of  
supersymmetry breaking and that through the boundary conditions 
of fields in supersymmetric quantum field theory on compactified 
space.
\section{Supersymmetric Quantum Mechanics on $S^1$}
We shall study effects of the quantization ambiguity, which implies 
that quantization on a space with nontrivial topology yields an 
undetermined parameter, on supersymmetric ground state wave functions of 
supersymmetric quantum mechanics on $S^1$. 
Let us consider a system in which there is the fermionic operator 
${\hat Q}_i$ that commutes with the Hamiltonian $\hat H$ and 
satisfies the supersymmetry algebra
\begin{equation}
[{\hat Q}_i, {\hat H} ]=0,\quad 
\{{\hat Q}_i, {\hat Q}_j\}=\delta_{ij} \hat H,
\qquad i=1,\cdots N.
\label{susy}
\end{equation} 
$N=2$ is the simplest case and it is of our interest. 
\par
Since the Hamiltonian is positive semidefinite, a supersymmetric 
state $\hq_i \ket{\Psi}=0$ is automatically a zero-energy ground state.
Conversely, if we have a zero-energy state, it has to be a supersymmetric 
ground state. Thanks to this property, finding supersymmetric ground states
is reduced to solving simple first-order equations
instead of solving the second-order equation ${\hat H}\ket{\Psi}=0$.
A key point for our study is that the fermionic 
operator $\hq_i (i=1, 2)$ should be written in terms of the operators 
which are appropriate
to describe the quantum mechanics on $S^1$ as shown in the 
subsection ${\bf 2.2}$. 
And we shall study the supersymmetric ground state wave functions 
of such a system \footnote{Concerning supersymmetric 
quantum mechanics and the ground state wave functions, it is known 
that a supersymmetric quantum mechanical system can be constructed by 
using the ground state 
wave functions of a nonsupersymmetric Hamiltonian \cite{gozzi}. 
In this approach the superpotential can be given by
the ground state 
wave function of the system. And the approach is extended to 
generally covariant systems such as relativistic particles interacting with 
external gauge fields and gravitational fields {\it etc}. \cite{gamboa}.}.
\par
\subsection{Quantum Mechanics on $S^1$}
Before we proceed to a model of supersymmetric quantum mechanics 
on $S^1$, it may be important and instructive to review briefly 
the Ohnuki-Kitakado formulation of quantum mechanics on $S^1$ \cite{ok}.
Those who are familiar with their formulation can skip this subsection and 
go directly to the subsection ${\bf 2.2}$ where the supersymmetric
quantum mechanics on $S^1$ is introduced. The discussions below are 
based on a paper \cite{tanimura} in a part of which the quantum mechanics 
on $S^1$ is summarized clearly. 
\par
The quantum mechanics on $S^1$ is 
defined by a self-adjoint operator $\hg$ and a 
unitary operator $\hat W$ satisfying the commutation relation 
\begin{equation}
[\hat G, \hat W]=\hbar~{\hat W}.
\label{comm}
\end{equation}
The operators $\hg, \hw$, and $\hwd$ generate an algebra.
Let us construct its representation. We shall start with 
an eigenvalue equation
\begin{equation}
{\hg} \ket{\alpha}=\hbar~\alpha \ket{\alpha}\qquad{\rm with}
\qquad \norm{\alpha}{\alpha}=1,
\end{equation}
where an eigenvalue $\alpha$ is a real number.
It is easy to see that $\hw(\hwd)$ raises (lowers) the eigenvalues of $\hg$
\begin{equation}
{\hg}{\hw}\ket{\alpha}=\hbar~(\alpha+1)\hw \ket{\alpha},\qquad
{\hg}{\hwd}\ket{\alpha}=\hbar~(\alpha-1)\hwd\ket{\alpha}.
\end{equation}
A state vector defined by
\begin{equation}
\ket{n+\alpha}\equiv {\hw}^n \ket{\alpha}, \qquad n={\rm integer},
\end{equation}   
is also an eigenstate of $\hg$:
\begin{equation}
\hg\ket{n+\alpha}=\hbar~(n+\alpha)\ket{n+\alpha}.
\label{irred}
\end{equation} 
For fixed $\alpha$, our Hilbert space, denoted 
by ${\cal H}_{\alpha}$ where the two operators $\hg, \hw$
are defined, is given by completing the vector space of linear combinations
of $\ket{n+\alpha} (n=0,\pm 1, \pm 2, \cdots)$.
The set of state vectors forms the orthocomplete system in 
${\cal H}_{\alpha}$. Therefore, we have 
\begin{equation}
\norm{m+\alpha}{n+\alpha}=\delta_{mn}, \quad
\sum_{n=-\infty}^{+\infty}\ket{n+\alpha}\bra{n+\alpha}={\bf 1}_{\alpha},
\label{irred2}
\end{equation}
where ${\bf 1}_{\alpha}$ is an identity operator in ${\cal H}_{\alpha}$.
Equation (\ref{irred}) and $\hw\ket{n+\alpha}=\ket{n+1+\alpha}$ define 
an irreducible representation of the algebra (\ref{comm}) 
on ${\cal H}_{\alpha}$. The classification of the irreducible 
representation of the algebra may be done 
by noting that (i) ${\cal H}_{\alpha}$ and 
${\cal H}_{\beta}$ are unitary equivalent 
Hilbert space if and only if $\alpha - \beta ={\rm integer}$ and (ii)
for an arbitrary irreducible representation $\cal H$ of the algebra, 
there exists a real number $\alpha$ such that $\cal H$ is 
the unitary equivalent of ${\cal H}_{\alpha}$.     
Thus, the classification is completed; that is, all the inequivalent 
irreducible representations 
are given by the Hilbert space ${\cal H}_{\alpha} (0 \leq \alpha < 1)$.
It should be emphasized that the algebra (\ref{comm}) 
has an infinite number of inequivalent representations characterized 
by an undetermined parameter $\alpha$, as contrary to the usual 
irreducible representation of the canonical algebra on 
one-dimensional Euclidean space.
\par
So far, we have constructed the $\hg$-diagonal representation. One can also 
go to the $\hw$-diagonal representation by which we will obtain wave functions
in the quantum mechanics on $S^1$. For fixed representation 
space ${\cal H}_{\alpha}$, since $\hw$ is a unitary operator, the eigenvalue 
equation for it may be written as 
\begin{equation}
\hw \ket{\theta}= \e^{i\theta}\ket{\theta}.
\label{eigen}
\end{equation}   
Its solution is
\begin{equation}
\ket{\theta} = \kappa(\theta)\sum_{n=-\infty}^{+\infty}
\e^{-i n \theta}\ket{n+\alpha},
\end{equation}
where $\theta$ is a real parameter and $\kappa(\theta)$ is an arbitrary
complex-valued function satisfying 
$\abs{\kappa(\theta)}=1$ and $\kappa(\theta+2\pi)=\kappa(\theta)$.
It is not difficult to show that 
\begin{eqnarray}
\ket{\theta+2\pi n}&=&\ket{\theta}, \qquad n={\rm integer},
\label{period}\\
\norm{\theta}{\theta^{\prime}}&=&2\pi \sum_{n=-\infty}^{n=+\infty}
\delta(\theta-\theta^{\prime}+2\pi n),
\label{ortho}\\
\int_0^{2\pi}{{d\theta}\over {2\pi}}\ket{\theta}\bra{\theta}&=&
\sum_{n=-\infty}^{+\infty}\ket{n+\alpha}\bra{n+\alpha}={\bf 1}_{\alpha},
\label{compl}\\
{\rm exp}(-i\lambda{\hg\over\hbar})\ket{\theta}&=&
\e^{-i\lambda\alpha}\kappa(\theta)\kappa^*(\theta+\lambda)
\ket{\theta+\lambda}, 
\label{trans}
\end{eqnarray}
where ${\bf 1}_{\alpha}$ is an identity operator in ${\cal H}_{\alpha}$.
These correspond to periodicity, orthonormality, completeness, and 
translation for the eigenstate of $\hw$. Let us note that 
it may be possible from Eqs. (\ref{eigen}) and (\ref{trans}) to 
identify $\hg$ and $\hw$ with the momentum and 
the position operators on $S^1$, respectively. 
\par
Now, let $\ket{\psi}$ be a state vector and we define a wave function
$\psi(\theta)$ on $S^1$ as follows
\begin{equation}
\psi(\theta)\equiv \norm{\theta}{\psi}.
\end{equation}
Taking the inner product of Eq. (\ref{trans}) with $\ket{\psi}$, we obtain
\begin{equation}
\bra{\theta}{\rm exp}(i\lambda{\hg\over\hbar})\ket{\psi}
=\e^{i\lambda\alpha}\kappa^*(\theta)\kappa(\theta+\lambda)
\norm{\theta+\lambda}{\psi}, 
\end{equation}
from which the $\hw$-diagonal representation of $\hg$ is given by
\begin{equation}
\bra{\theta}{\hg}\ket{\psi}=
\Bigl[-i\hbar~{\del\over{\del\theta}}-i\hbar~\kappa^*(\theta)
{{\del\kappa(\theta)}\over {\del\theta}}+\hbar~\alpha\Bigr]\psi(\theta).
\label{repreg}
\end{equation}
We also obtain, from Eq. (\ref{eigen}),
\begin{equation}
\bra{\theta}{\hw}\ket{\psi}=\e^{i\theta}\psi(\theta).
\label{reprew}
\end{equation}
The inner product on $S^1$ is expressed in terms 
of the wave function as
\begin{equation}
\norm{\chi}{\psi}=\int_0^{2\pi}{{d\theta}\over {2\pi}}
\chi^*(\theta)\psi(\theta).
\end{equation} 
Thus, the representation of Hilbert space, which is defined by
Eqs. (\ref{repreg}) and (\ref{reprew}), is the space of the square integrable 
function on $S^1$. Let us note that all wave functions 
have to satisfy the periodic boundary condition $\psi(\theta+2\pi n)
=\psi(\theta)$, which is a direct consequence of Eq. (\ref{period}). 
This periodicity is essential when we study the supersymmetric ground state
wave functions of the supersymmetric quantum mechanics on $S^1$.
\par
Let us next present the physical meaning of the parameter $\alpha$. 
To this end, let us redefine $\kappa(\theta)$ 
by utilizing the arbitrariness of it in 
such a way that $\kappa(\theta)=\omega(\theta)\kappa^{\prime}(\theta)$, where 
$\omega(\theta)$ has to satisfy $\abs{\omega(\theta)}=1$ 
and $\omega(\theta+2\pi)=\omega(\theta)$. It follows that    
$\ket{\theta}=\omega(\theta)\ket{\theta}^{\prime}$, so that the transformed
wave function $\psi^{\prime}(\theta)$ is given by
\begin{equation}
\psi^{\prime}(\theta)=\omega(\theta)\psi(\theta).
\label{gtrf1}
\end{equation} 
According to this redefinition, the $\hw$-diagonal representation for $\hg$
becomes
\begin{equation} 
'\bra{\theta}{\hg}\ket{\psi}=
\Bigl[-i\hbar~{\del\over{\del\theta}}+A^{\prime}(\theta)\Bigr]
\psi(\theta),
\label{newrepreg}
\end{equation} 
where we have defined 
\begin{equation}
A^{\prime}(\theta) \equiv A(\theta)+i\hbar~\omega^*(\theta)
{{\del\omega(\theta)}\over{\del\theta}}, \quad
A(\theta)\equiv -i\hbar~\kappa^*({\theta})
{{\del\kappa(\theta)}\over{\del\theta}}+\hbar~\alpha.
\label{gtrf2}
\end{equation}
Equations (\ref{gtrf1}) and (\ref{gtrf2}) stand for the gauge transformation.
Therefore, the parameter $\alpha$ has the meaning of the gauge field.
It is easy to see that the gauge field has 
the properties (i) $A(\theta)$, assumed to be   
an arbitrary real-valued function satisfying the periodic boundary condition 
$A(\theta+2\pi)=A(\theta)$, can always be made a constant 
function $A^{\prime}(\theta)=\alpha$ by a gauge transformation and (ii) for 
two constant functions $A^{\prime}(\theta)=\alpha$ 
and $A^{\prime}(\theta)=\beta$, these are 
connected by a unique gauge transformation if and only if $\beta-\alpha$ is
an integer. Thus, we arrive at an important conclusion that 
all the inequivalent gauge fields are given by 
$A_{\alpha}\equiv \alpha (0\leq \alpha < 1)$\footnote{
Dirac's approach to the quantization for a constrained system does not
yield an infinite number of inequivalent representations; that is, it 
corresponds to only $\alpha=0$. Quantization and 
embedding $S^1$ into higher dimensional space ${\bf R}^2$ is 
not a ``commutative'' procedure relating to each other. Let us also note that 
the Ohnuki-Kitakado formulation is independent of the dynamics, in contrast to 
Dirac's approach.}.
Hereafter, we choose $\kappa(\theta)=1$ for simplicity.
\par
It is a very special feature of the quantum mechanics on $S^1$ that the 
inequivalent gauge field is restricted to be $0\leq \alpha <1$. 
Another way of looking at it is that
if we perform a gauge transformation by
$
\psi(\theta)\rightarrow \psi^{\prime}(\theta)=\e^{in\theta}\psi(\theta),
$
we see that the gauge fields $A(\theta)$ and $A(\theta)-n\hbar$ 
are equivalent for $n=$integer. $n$ has to be restricted to be 
an integer; otherwise, the transformed wave 
function $\psi^{\prime}(\theta)$ does not satisfy 
the required periodic boundary condition. 
Therefore, the inequivalent gauge field 
is given by $0\leq A(\theta) < \hbar$~, which means $0\leq \alpha <1$. 
Let us note that the gauge transformation by $\e^{in\theta}$ with
$n=$ noninteger is a singular gauge transformation and is strictly
forbidden. 
\par
Different values of the gauge field give different 
quantum mechanics on $S^1$. It may be helpful to note that the 
gauge field $\alpha$ may correspond to   
the magnitude of the magnetic flux $e\Phi/2\pi \hbar~c$ through $S^1$ in 
the Aharanov-Bohm effect. The different magnitude of the flux 
actually gives different physics.   
\par
\subsection{Supersymmetric Simple Pendulum}
Now, we are ready to introduce a model of the supersymmetric quantum 
mechanics on $S^1$.
According the discussion above, the two 
operators $\hg$ and $\hw$, which correspond to the momentum 
and the position of a particle on
$S^1$, are fundamental. It may be natural to construct a quantum 
Hamiltonian in terms of these operators. The Hamiltonian in our model
is assumed to satisfy the supersymemtry algebra (\ref{susy}), so that 
the fermionic operators ${\hq}_i(i=1, 2)$ also have to be given in 
terms of them.
We will discuss the classical counterpart of the quantum hamiltonian 
constructed in this way later.
\par
Let us define the fermionic operator ${\hq}_i(i=1,2)$ by
\begin{eqnarray}
\hq&\equiv&{1\over{\sqrt 2}}\Bigl({\hat Q}_1+i{\hat Q}_2\Bigr) \nonumber\\
&=&\Bigl({1\over{\sqrt{2m}R}}\hg +i V(\hw,\hwd)\Bigr)\hxi
\equiv {\hat q}~\hxi,
\label{supercharge}\\
\hqb&\equiv&{1\over{\sqrt 2}}\Bigl({\hat Q}_1-i{\hat Q}_2\Bigr)
\nonumber\\
&=&\Bigl({1\over{\sqrt{2m}R}}\hg-iV(\hw,\hwd)\Bigr)\hxib
\equiv {\hat q}^{\dagger}~\hxib.
\label{supercharge1}
\end{eqnarray}
$V(\hw,\hwd)$, which is called the superpotential hereafter, is a hermitian 
operator in terms of $\hw$ and $\hwd$. Here $m$ and $R$ stand for 
the mass of a particle and the radius of $S^1$, respectively.
The fermionic variables $\hxi, \hxib$ satisfy the algebra
\begin{equation}
\{\hxi,\hxib\}=1,\qquad \hxi^2=\hxib^2=0.
\label{anticomm}
\end{equation}
Then, the Hamiltonian is given by
\begin{eqnarray}
\hat H &=& \{{\hat Q},{\hat {\bar Q}}\} \nonumber \\
&=& {1\over{2m R^2}}\hg^2+V^2(\hw,\hwd)\nonumber\\
&-&{i\over {{\sqrt{2m}R}}}
\Bigl(\hg V(\hw,\hwd)-V(\hw,\hwd)\hg\Bigr)\Bigl[\hxi, \hxib\Bigr],
\label{hamilton1}
\end{eqnarray}
where we have used Eq. (\ref{anticomm}).  
\par
Here, it may be necessary to discuss the classical counterpart 
of the quantum Hamiltonian (\ref{hamilton1}). To this end, let us note that
the fundamental algebra (\ref{comm}) may be actually inferred 
by the classical Poisson's brackets for the angle variable $\theta$ and
the correponding momentum $P_{\theta}$ in the polar coordinate:
\begin{equation}
\{P_{\theta}, \e^{i\theta}\}_P=-i \e^{i\theta}.
\end{equation}
If we replace the 
classical Poisson's brackets by the commutation relation 
divided by $i\hbar$~, we obtain
the fundamental algebra (\ref{comm}) by identifying $\e^{i\theta}$ 
and $P_{\theta}$ 
with $\hw$ and $\hg$, respectively. This is the same identification
stated earlier. Therefore, in the classical limit we may replace
$\hw$ by $\e^{i\theta}$ and $\hg$ by $P_{\theta}$.
According to these replacements, we obtain a classical 
Hamiltonian, ignoring the fermionic variables $\hxi, \hxib$:
\begin{equation}
\hat H\rightarrow H_{cl}={P_{\theta}^2\over {2mR^2}}+
V^2(\e^{i\theta},\e^{-i\theta}).
\end{equation}
If we choose $V(\e^{i\theta},\e^{-i\theta})$ as
\begin{equation}
V(\e^{i\theta},\e^{-i\theta})=\sqrt{{{mg_NR}\over 2}}\sin\theta,
\label{poten1}
\end{equation}
the classical Hamiltonian describes a simple pendulum with angle 
$2\theta$. Here $g_N$ is the gravitation accelerator constant.
\par
On the other hand, given the superpotential (\ref{poten1}), the quantum
counterpart of it is obtained by 
\begin{equation}  
V(\hw,\hwd)=\sqrt{{mg_NR}\over {2}}\Bigl({{\hw-\hwd}\over {2i}}\Bigr).
\end{equation}
Having this superpotential, the model of the supersymmetric quantum
mechanics on $S^1$, that is, the {\it supersymmetric simple pendulum}, is 
given by the Hamiltonian
\begin{equation}
\hat H= {1\over {2mR^2}}\hg^2+{{mg_NR}\over 2}\sin^2\theta 
-{\hbar\over 2}\sqrt{{g_N\over R}}\cos\theta\Bigl[\hxi, \hxib\Bigr]
\label{hamilton2} 
\end{equation}
in the $\hw$-diagonal representation. 
It is understood that $\hg=-i\hbar~\del/\del\theta+\hbar~\alpha$ and 
$\alpha$ is the gauge field discussed in the previous subsection.
\par
Let us study the supersymmetric ground state wave functions of 
the supersymmetric simple pendulum whose hamiltonian 
is given by Eq. (\ref{hamilton2}). 
It follows from the algebra (\ref{susy}) that
the supersymmetric ground states must be zero-energy states 
satisfied by
\begin{equation}
\hq \ket{\Psi}=0\qquad{\rm and}\qquad  \hqb\ket{\Psi}=0.
\label{ground1}
\end{equation}
Let us introduce a matrix representation for the fermionic variables.
It is easy to see that the matrix representations given by 
\begin{equation}
\hxi=\left(\begin{array}{cc}
0&0\\
1&0\end{array}\right),\qquad
\hxib=\left(\begin{array}{cc}
0&1\\
0&0\end{array}\right)
\label{supercharge2}
\end{equation}
satisfy the algebra (\ref{anticomm}). Then, it follows that
$$
\Bigl[\hxi, \hxib \Bigr]=-\left(\begin{array}{cc}
1&0\\
0&-1\end{array}\right)\equiv -\sigma^3.
$$
Using these, the 
Hamiltonian (\ref{hamilton2}) becomes
\begin{eqnarray}
\hat H&=& \Bigl({1\over {2mR^2}}\hg^2+{{mg_NR}\over 2}\sin^2\theta\Bigr)
{\bf 1}_{2\times 2}+{\hbar\over 2}\sqrt{{g_N\over R}}\sigma^3\cos\theta
\nonumber\\
&=&\left(\begin{array}{cc}
{\hat q}^{\dagger}{\hat q}&0\\
0&{\hat q}{\hat q}^{\dagger}\end{array}\right)\equiv 
\left(\begin{array}{cc}
{\hat H}_{+}&0\\
0&{\hat H}_{-}\end{array}\right).
\label{hamilton3} 
\end{eqnarray}
In the matrix representation, the 
hamiltonian is a $2\times 2$ matrix. 
\par
Since the Hamiltonian (\ref{hamilton3}) commutes with an  
operator ${\hat S}^{F}\equiv \sigma^3/2$, the eigenstates 
of the Hamiltonian is labeled by the eigenvalues of ${\hat S}^F$. 
Let us call the two states
$\ket{+}$ and $\ket{-}$ fermion 
numbers $+1/2$ and $-1/2$, respectively\footnote{One may call 
the two states spin-up and spin-down states. Or by 
introducing ${\hat f}=\half+\half[\hxi, \hxib]$
whose eigenvalues are $0, 1$, one may denote them by fermion 
numbers $0, 1$.}. 
The state vector is, now, a two-component vector
\begin{equation}
\ket{\Psi}={\ket{+}\choose \ket{-}}.
\label{spinor}
\end{equation}
In the $\hw$-diagonal representation, it may be written as
\begin{equation}
\Psi(\theta)={\psi_{+\half}(\theta)\choose \psi_{-\half}(\theta)}.
\end{equation}
The Hamiltonian ${\hat H}$ is diagonalized with respect to the 
fermion number $\pm 1/2$. In this matrix 
representation, Eq. (\ref{ground1}) is read as
\begin{eqnarray}
{\hat q}~\psi_{+\half}(\theta)=
\Bigl({1\over {\sqrt{2m}R}}(-i\hbar~{\del\over{\del\theta}}+\hbar~\alpha) 
+ i\sqrt{{mg_NR\over 2}}\sin\theta\Bigr)~\psi_{+\half}(\theta)=0,
\nonumber\\
{\hat q}^{\dagger}\psi_{-\half}(\theta)=
\Bigl({1\over {\sqrt{2m}R}}(-i\hbar~{\del\over{\del\theta}}+\hbar~\alpha) 
- i\sqrt{{mg_NR\over 2}}\sin\theta\Bigr)~\psi_{-\half}(\theta)=0.
\label{ground2}
\end{eqnarray} 
Solutions for Eq. (\ref{ground2}) are found to be 
\begin{equation}
\psi_{+\half}(\theta)
={1\over \sqrt{I_0(2z)}}{\rm exp}(-i\alpha\theta-{z\over {\hbar}}\cos\theta),
\quad
\psi_{-\half}(\theta)
={1\over \sqrt{I_0(2z)}}{\rm exp}(-i\alpha\theta+{z\over{\hbar}}\cos\theta),
\label{ground3}
\end{equation} 
where $I_0(2z)$ in the normalization factor is the zeroth-order modified
Bessel function and we have defined a dimensionless parameter
$$
{{mR^2}\over \hbar}\sqrt{{g_N\over R}}
\equiv {mR^2\over \hbar}\omega
\equiv {z\over {\hbar}}~~.
$$
\par
These are normalizable solutions. Thus, one may 
say that the zero-energy states, that is, supersymmetric ground 
states, exist in the model and the supersymmetry is unbroken. 
This is, however, a hasty conclusion. 
In addition to the normalizability, all the wave functions 
have to satisfy the periodic boundary condition
$\Psi(\theta+2\pi)=\Psi(\theta)$, that is, $\psi_{\pm \half}(\theta+2\pi)
=\psi_{\pm \half}(\theta)$, which follows 
from Eq. (\ref{period}) \footnote{Let us note that the two-component spinor 
state (\ref{spinor}) does not have a minus  
sign under $2\pi$ rotation in this case because the 
rotation is done by the usual rotation matrix in two dimensions.}. 
It is easy to see from Eq. (\ref{ground3}) that
\begin{equation}
\psi_{\pm \half}(\theta+2\pi)=\e^{-i2\pi\alpha}\psi_{\pm \half}(\theta).
\label{boundary}
\end{equation}
The boundary condition for the zero-energy state 
wave functions is 
twisted by the gauge field $\alpha$.  
The zero-energy state wave functions do not satisfy the required periodic 
boundary condition except for $\alpha={\rm integer}$. Since the inequivalent
representation is given by $0\leq \alpha <1$, they 
are inconsistent with the periodic boundary condition and become 
{\it unphysical} wave functions for $0< \alpha < 1$.
Therefore, the supersymmetry can be broken due to the 
gauge field $\alpha$. Let us note that the 
Witten index ${\rm Tr}(-1)^{{\hat f}}=n_B^{E=0}-n_F^{E=0}$ vanishes 
in our model.  
It is easy to see that $n_B^{E=0}=n_F^{E=0}=1$ for $\alpha={\rm integer}$
and that $n_B^{E=0}=n_F^{E=0}=0$ for $\alpha=$ noninteger.
\par
Unlike the usual supersymmetry breaking, in which the leading term of the
superpotential determines whether or not supersymmetry is broken, our
breaking of supersymmetry does not 
depend on the structure of the superpotential. It is entirely 
due to the existence of the gauge field $\alpha$, which 
is an inevitable consequence of the quantization ambiguity 
when one quantizes the theory on topologically nontrivial 
space such as $S^1$. The gauge field has the effect of twisting the boundary 
conditions of the zero-energy state wave functions. 
Among the various supersymmetric quantum 
mechanics on $S^1$ led by the gauge field, it includes theories 
with broken supersymmetry due to the gauge field.
Let us note that in this context there is no mechanism to determine
the values of $\alpha$ or what values of $\alpha$ we should take. 
\par
Note that for noninteger values of the gauge field $\alpha$, the gauge 
field can not be 
removed in Eq. (\ref{ground2}) by the gauge transformation  
\begin{equation}
\psi_{\pm \half}(\theta)\longrightarrow 
\psi_{\pm \half}^{\prime}(\theta)=\e^{-i\alpha\theta}\psi_{\pm \half}(\theta).
\label{gaugetrf}
\end{equation}
As we stated before, the inequivalent gauge field is given by
$0\leq \alpha < 1$. Any gauge field in this range cannot be connected by
a regular gauge transformation with $\e^{in\theta}(n={\rm integer})$. 
Only a singular gauge transformation can do it, but it destroys the required
periodic boundary condition for the transformed wave function.      
The singular gauge transformation is strictly 
forbidden, so that the gauge field cannot be gauged away.
\par
Let us briefly comment on the ground state energy. We have a physical
supersymmetric ground state wave function for $\alpha={\rm integer}$, so that 
the ground state energy is exactly zero. 
On the other hand, for $0< \alpha < 1$, supersymmetry 
is broken. The ground state energy is nonzero (positive).
Estimating the ground state energy is a subtle 
problem as studied in \cite{abbott1}\cite{abbott2}. 
\par
The Hamiltonian (\ref{hamilton3}) cannot be solved 
analytically. The bosonic potential ${{mg_N R}\over 2}\sin^2\theta$
is periodic and the classical vacuum has a periodic structure. One may expect 
that there is a instantonlike classical solution, which gives a finite 
Euclidean action, connecting the two vacua with different 
fermion numbers $(\pm 1/2)$. Actually, there exists such a classical 
solution in our model. It is given by 
$
\cos\theta_{cl}(\tau)=\pm \tanh(\omega(\tau-\tau_0))
$
with the classical Euclidean action being $-2z/\hbar$~. And the fermion zero 
mode exists in this classical background. Therefore, we expect 
tunneling to occur between the two vacua. According to the semiclassical 
argument, the tunneling effect shifts the ground state energy to give 
an exponentially small amount of energy in the form of
${\rm exp}(-2z/\hbar)\times \cos{2\pi\alpha}$\footnote{This band structure
$\cos{2\pi\alpha}$ can be understood from the effective action 
obtained by the transition amplitude $K(\theta_f,t;\theta_i,0)
=\bra{\theta_f}{\exp(-i{\hat H}t/\hbar)\ket{\theta_i}}=
\sum_{n=-\infty}^{+\infty}\int_{n-winding} {\cal D}\theta 
\exp(iS_{eff}/\hbar)$, where the effective action is given by 
$S_{eff}=\int dt {{mR^2}\over 2}({{d\theta}\over {dt}})^2
-{{mg_NR}\over 2}\sin^2\theta
+\half\sqrt{g_N\over R}\cos\theta[\xi, \xi^*]+i\xi^*{{d\xi}\over {dt}}
-\alpha{d\theta\over {dt}}$. The ``topological'' term 
$\alpha{\dot\theta}$ is the origin of such band structure\cite{raja}.}. 
\par
On the other hand, for very small $z$, we
can resort to perturbation theory to obtain the energy spectrum of
the Hamiltonian. The ground state energy is given by 
$E_0\sim {1\over {2mR^2}}(\alpha^2+O(z^2))$, where we have set $\hbar=1$. 
The gauge field $\alpha$ is a dominant contribution to the ground state
energy in this case.    
\par
It may be interesting to consider the $R\rightarrow \infty$ limit.
So far, we have fixed the radius $R$ of $S^1$. If $R$ varies to become 
large, we expect that the arc of an arbitrary part of $S^1$ will approach 
a straight segment. In the limit of $R\rightarrow \infty$, one-dimensional 
Euclidean space will be recovered.\par
In order to study the limit, let us define a variable $x\equiv R\theta$.
In terms of this new variable, the Hamiltonian (\ref{hamilton3}) is written 
as, remembering $\hg=-i\hbar~\del/\del\theta +\hbar~\alpha$ in the 
$\hw$-diagonal representation, 
\begin{equation}  
\hat H= \Bigl({-\hbar^2\over {2m}}\Bigl({\del\over {\del x}}
+i{\alpha\over R}\Bigr)^2
+{{mg_NR}\over 2}\sin^2({x\over R})\Bigr){\bf 1}_{2\times 2} 
+{\hbar\over 2}\sqrt{{g_N\over R}}\sigma^3\cos({x\over R}).
\label{hamilton4} 
\end{equation}
If we take the limit of $R\rightarrow\infty$ naively, it becomes trivial for 
the Hamiltonian to yield the one for a free particle on $S^1$:
\begin{equation}
{\hat H}={-\hbar^2\over {2m}}{\del^2\over {\del x^2}}{\bf 1}_{2\times 2}.
\end{equation} 
In order to obtain an interacting theory, one has to take the limit, keeping 
a relation given by
\begin{equation}
{m\over \hbar}\sqrt{g_N\over R}\equiv 
{{m\omega}\over \hbar}=\Bigl({\rm strength~of~oscillator} \Bigr)^2
={\rm const}.
\label{relation}
\end{equation}
Then, we obtain 
\begin{equation}  
\hat H = \Bigl({-{\hbar}^2\over{2m}}{\del^2\over {\del x^2}}
+{{m\omega^2}\over 2} x^2\Bigr){\bf 1}_{2\times 2} 
+{{\hbar~\omega}\over 2}\sigma^3 + O({1\over R^2}).
\label{harmo} 
\end{equation}
This is the well-known Hamiltonian for the supersymmetric harmonic 
oscillator \cite{witten} with angular frequency $\omega=\sqrt{g_N/R}$.
Likewise, by taking the same limit, the fermionic operators $\hq, \hqb$
become
\begin{eqnarray}
\hq&=&{1\over {\sqrt {2m}}}\Bigl({\hat p} + i W({\hat x})\Bigr)\hxi
+O({1\over R^2})\equiv \hq_{susy}+O({1\over R^2}),\nonumber\\
\hqb&=&{1\over {\sqrt{2m}}}\Bigl({\hat p} - i W({\hat x})\Bigr)\hxib
+O({1\over R^2})\equiv\hqb_{susy}+O({1\over R^2}),
\label{super2}
\end{eqnarray}
where we have defined $W({\hat x})\equiv m{\omega} {\hat x}$ and 
${\hat p}\equiv -i{\hbar}~\del/\del x$. It is easy to check that 
the Hamiltonian (\ref{harmo}) satisfies the supersymmetry algebra
(\ref{susy}) with the supercharges (\ref{super2}) if we use the 
canonical commutation relation $[{\hat p},{\hat x}]=-i\hbar$~~and 
Eq. (\ref{supercharge2}). The usual supersymmetry, by which the supersymmetry 
transformations between boson $({\hat x})$ and fermion $(\hxi, \hxib)$ 
are generated, is realized 
in the limit of $R\rightarrow \infty$ with Eq. (\ref{relation}). 
\par
The supersymmetric ground state wave functions for the Hamiltonian
(\ref{harmo}) are obtained by solving the first-order equation
$\hq_{susy}\ket{\Psi}=0$ and $\hqb_{susy}\ket{\Psi}=0$. Using the
same matrix representation as before, their solutions are
\begin{equation}
{\psi}_{+\half}^{h.o}(x)\sim {\rm exp}(+{{m\omega}\over {2\hbar}} x^2),
\qquad
{\psi}_{-\half}^{h.o}(x)= \Bigl({{m\omega}\over {\pi\hbar}}\Bigr)^{1/4}
{\rm exp}(-{{m\omega}\over {2\hbar}} x^2). 
\label{harmosol}
\end{equation}
There are two candidates for the supersymmetric ground 
state (zero-energy state) wave functions. 
The one is physical and its wave function is given 
by ${\psi}_{-\half}^{h.o}(x)$. The other 
one ${\psi}_{+\half}^{h.o}(x)$ is unphysical because of its 
non-normalizability. Therefore, we have one supersymmetric 
ground (zero-energy) state. This is consistent with the exact energy 
spectrum of the Hamiltonian (\ref{harmo}). As easily seen from the 
Hamiltonian, there exists one supersymmetric ground state.
In fact, the solutions (\ref{harmosol}) can be obtained by taking the 
the limit of $R\rightarrow \infty$ with the relation (\ref{relation}) 
in Eq. (\ref{ground3}). 
By noting $I_0(z)\sim \e^z/\sqrt{2\pi z}$ for large $z$,
we obtain
\begin{equation}
{\tilde\psi}_{+\half}\sim {\rm exp}(+{{m\omega}\over {2\hbar}} x^2),\qquad
{\tilde\psi}_{-\half}(x)= \Bigl({{m\omega}\over {\pi\hbar}}\Bigr)^{1/4}
{\rm exp}(-{{m\omega}\over {2\hbar}} x^2),
\end{equation}
where we have redefined the normalization as 
${\tilde\psi}_{\pm \half}(x)dx\equiv 
\psi_{\pm \half}(\theta){{d\theta}\over{\sqrt{2\pi R}}}$. 
These are the same as Eqs. ({\ref{harmosol}).
The Witten index is ${\rm Tr}(-1)^{{\hat f}}=1$ in this case. 
\par
A simple generalization of the model is an $N$-component one. 
The fermionic operators are defined by
\begin{eqnarray}
\hq &=&\sum_{a=1}^{N}
\Bigl({1\over{\sqrt{2m}R_a}}\hg_a +i {\hat V}_a\Bigr)\hxi_a \equiv 
\sum_{a=1}^{N}{\hat q}_a~\hxi_a,
\label{nsupercharge}\\
\hqb&=&\sum_{a=1}^{N}
\Bigl({1\over{\sqrt{2m}R_a}}\hg_a-i{\hat V}_a\Bigr)\hxib \equiv 
\sum_{a=1}^{N}{\hat q}_a^{\dagger}~\hxib_a,
\label{nsupercharge1}
\end{eqnarray}
where ${\hat V}_a\equiv {\hat V}_a(\theta_1, \cdots, \theta_N)$ and 
$\hg_a=-i\hbar~\del/\del\theta_a+\hbar~\alpha_a$ in 
the $\hw$-diagonal representation. Let us assume 
\begin{equation}
[\hg_a, \hw_b]=\hbar~\delta_{ab}\hw_b, \quad
\{\hxi_a, \hxib_b\}=\delta_{ab}, \quad
\{\hxi_a, \hxi_b\}=0, \quad
\{\hxib_a, \hxib_b\}=0. \quad
\end{equation}
Then, the Hamiltonian following from these fermionic operators is
\begin{equation}
{\hat H}=\sum_{a=1}^N{1\over {2 m R_a^2}}\hg_a\hg_a + {\hat V}_a{\hat V}_a
-\sum_{a, b=1}^N{i\over {{\sqrt{2m}R_a}}}
[\hg_a, {\hat V}_b][\hxi_a, \hxib_b].
\label{torus}
\end{equation}
The Hamiltonian (\ref{torus}) may describe the supersymmetric quantum 
mechanics on the torus $T^N=S^1\otimes\cdots\otimes S^1$. 
As before, $\alpha_a (a=1, \cdots, N)$ may be interpreted as the gauge field 
appearing as a consequence of the quantization on each topological 
space $S^1$. 
\par
It is difficult to obtain the exact form of the supersymmetric 
ground state wave functions of the model. If we, however, restrict 
ourselves to
certain sectors of the model, they can be obtained in closed 
form such as Eq. (\ref{closed}) \cite{claud}. 
In order to see this, let us define 
\begin{equation}
\ket{-}\equiv \ket{0},\quad \ket{+}\equiv \prod_{a=1}^N\hxib_a~\ket{0}\quad
{\rm with}\quad \hxi_a \ket{0}=0~~(a=1, \cdots, N).
\end{equation}
Then, $\hq\ket{-}=\hqb\ket{+}=0$ is trivially satisfied, so that in these 
two sectors, the supersymmetric ground state (zero-energy state) wave 
functions are obtained in closed form by solving simple first-order
equations such as Eqs. (\ref{ground2}). 
Aside from the normalization, the solutions 
are obtained as
\begin{eqnarray}
\Psi^{+}(\theta_1,\cdots, \theta_N)&=&
{\rm exp}\sum_{a=1}^N\Bigl(-i\alpha_a\theta_a 
+{{\sqrt{2m}R_a}\over \hbar}\int^{\theta_a}d{\bar\theta}_a~
V_a({\bar\theta}_1,\cdots, {\bar\theta}_a, \cdots, {\bar\theta}_N)\Bigr)
\ket{+},
\nonumber\\
\Psi^{-}(\theta_1, \cdots, \theta_N)&=&
{\rm exp}\sum_{a=1}^N\Bigl(-i\alpha_a\theta_a 
-{{\sqrt{2m}R_a}\over \hbar}\int^{\theta_a}d{\bar\theta}_a~
V_a({\bar\theta}_1,\cdots, {\bar\theta}_a, \cdots, {\bar\theta}_N)\Bigr)
\ket{-}.
\label{extend}
\end{eqnarray} 
The wave functions have to satisfy the periodic boundary condition
$\Psi^{\pm}(\cdots, \theta_a+2\pi, \cdots)
=\Psi^{\pm}(\cdots, \theta_a, \cdots)~(a=1,\cdots, N)$.
If the contributions coming from the superpotential in Eqs. (\ref{extend}) 
do not spoil the normalizablity and the periodicity of 
the wave functions, the supersymmetry can be broken
for noninteger values of $\alpha_a (a=1, \cdots, N)$.  
\par
\section{Conclusions and Discussion}
We have applied the Ohknuki-Kitakado formulation of quantum 
mechanics on $S^1$ to the supersymmetric 
simple pendulum whose Hamiltonian is given by Eq. (\ref{hamilton3}) and 
satisfies the algebra (\ref{susy}). 
According to their formulation, an undetermined parameter, which can be 
interpreted as a constant gauge field, inevitably
enters into the theory to yield the various quantum mechanics on $S^1$.  
We have studied the effects of the quantization ambiguity on the 
supersymmetric ground state wave functions of the model.
\par
We have found that supersymmetry can be broken due to the 
existence of the gauge field $\alpha$. The gauge field twists 
the boundary condition
of the supersymmetric ground state wave functions. 
For noninteger values of $\alpha$, they 
do not satisfy the required periodic boundary condition.
As a result, they become unphysical wave functions though they are 
normalizable. The mechanism of supersymmetry breaking 
is different from the usual supersymmetry breaking
discussed in supersymmetric quantum mechanics. 
The latter depends crucially on the structure, the leading term, of 
the superpotential, while the former is entirely due to the quantization 
ambiguity resulting firmly from 
quantization on a space with nontrivial topology like $S^1$. 
\par
We have chosen the superpotential $V(\hw,\hwd)$
in such a way that it becomes a simple pendulum in the classical limit.
In principle, one can choose any superpotential as long as it can be written 
in terms of integer powers of the operators $\hw$ and $\hwd$. 
Thanks to the factorizable property for finding  
the supersymmetric ground state wave functions,
they are given simply by solving the first order equation (\ref{ground1})
and are obtained in closed form
\begin{equation}
\psi_{\pm \half}({\theta})={\rm exp}\Bigl(-i\alpha\theta \mp 
{{\sqrt{2m}R}\over\hbar}\int^{\theta}d{\bar\theta}~V(\e^{i{\bar\theta}},
\e^{-i{\bar\theta}}) \Bigr),
\label{closed}
\end{equation} 
for the general superpotential.
Our mechanism of supersymmetry breaking is not altered by the 
choice of the superpotential
if ${\rm exp}(\int^{\theta}d{\bar\theta}
V(\e^{i{\bar\theta}},\e^{-i{\bar\theta}}))$ 
does not violate the periodicity and the 
normalizability of the wave functions, which is the case 
for the superpotential satisfying our criterion. 
Supersymmetry breaking will
always occur for noninteger values of $\alpha$. 
Because of the factorizable property, there is no way to prevent 
the gauge field from entering into the supersymmetric ground state wave 
functions and twisting their boundary conditions. 
The gauge field cannot be removed by a regular gauge transformation.
\par
One may wonder whether all the eigenfunctions of the 
Hamiltonian (\ref{hamilton3}) become unphysical, that is, those that do not
satisfy the periodic boundary condition due to the existence of the 
gauge field. This is not true. In order to see this, 
let us consider a free Hamiltonian, ignoring all terms except for 
$\hg^2$. The energy eigenvalue depends on $\alpha$ like $(m+\alpha)^2$ and
the gauge field produces an effect on the observable at 
the quantum level \cite{tanimura}. 
The corresponding eigenfunction satisfying the periodic boundary condition 
is easily found to be $\e^{im\theta}$. 
The ground state wave functions and the other eigenfunctions 
are obtained by solving
essentially different types of differential equations in 
the system satisfying the supersymmetry algebra (\ref{susy}).
\par
We have also discussed the limit of $R\rightarrow \infty$. 
One-dimensional Euclidean space is realized.
In the limit with the relation (\ref{relation}), we have obtained 
the supersymmetric 
harmonic oscillator with angular frequency $\omega=\sqrt{g_N/R}$. 
There exists one physical supersymmetric ground state wave function. 
The other one, though it is a zero-energy state, is unphysical because of 
its non-normalizability. These two wave functions are actually 
obtained by taking the limit in the solutions of the zero-energy wave 
functions (\ref{ground3}).
In the limit all the effects of the gauge field $\alpha$ 
disappear. Then, an infinite number of inequivalent representations 
is reduced to a unique representation, which is nothing but the 
representation of the canonical algebra $[{\hat p}, {\hat x}]=-i\hbar$~. 
The fermionic operators become supercharges in the same limit, and they
generate supersymmetry transformations between 
bosons $({\hat x})$ and fermions $(\hxi, \hxib)$.
\par
We have also considered the $N$-component generalization 
of Eq. (\ref{hamilton1})
and studied the supersymmetric ground state wave functions of the model.
The Hamiltonian (\ref{torus}) describes the supersymmetric quantum mechanics 
on the torus $T^N$. If we restrict ourselves to the 
two sectors given by $\ket{-}$ and $\ket{+}$, the wave functions 
can be obtained in closed form (\ref{extend}) by solving the simple 
first-order equations. We have found, again, that supersymmetry 
was able to be broken due to the existence of the gauge 
field $\alpha_a \neq {\rm integer}(a=1, \cdots, N)$, which appeared  
as a consequence of the quantization on the each topological space $S^1$. 
\par
Finally, let us discuss the similarities between our mechanism of 
supersymmetry breaking and that through boundary conditions
of fields for compactified directions in 
supersymmetric quantum field theory.
\par
Strictly speaking, supersymmetry breaking through boundary conditions
is one thing, and that through our mechanism is another.
Nevertheless, it may be interesting to discuss the similarities
between the two supersymmetry breakings. 
In the former case, the breaking means that the action is no 
longer invariant under the supersymmetry transformations. But it does 
not necessarily mean the nonexistence of the zero-energy state 
in the system. The supersymmetric 
harmonic oscillator at finite temperature is one of the examples in which
the action is not invariant under the supersymmetry transformations 
because of the different boundary conditions between the bosons and the 
fermions; there exists, however, the zero-energy state in the system, which 
results from ${\rm Tr}(-1)^{{\hat f}}=1$ \cite{fuchs}. 
In the latter case, we  assume  
that the Hamiltonian satisfies the supersymmetry algebra ({\ref{susy}), so 
that supersymmetry breaking immediately
means the nonexistence of physical zero-energy states 
in the system, and whether 
or not supersymmetry is broken is determined definitely
by the existence of the zero-energy state. 
\par
As seen from Eq. (\ref{boundary}), the boundary condition for the $S^1$ 
direction is twisted for $0 < \alpha < 1$ by $\e^{-i2\pi\alpha}$. 
If we consider a theory at finite temperature, it is equivalent to studying
the theory in a space where the Euclidean time direction 
is compactified on $S^1$. 
It is well known that supersymmetry is broken at finite
temperature by the different boundary conditions 
for the Euclidean time direction between the bosons and the fermions. 
The boson (fermion) satisfies 
the (anti)periodic boundary condition. The case of $\alpha=1/2$, which 
actually corresponds to the antiperiodic boundary condition, is similar
to the case of supersymmetry breaking at finite temperature. 
\par  
More generally, if one wishes to break 
supersymmetry through different boundary conditions between the 
bosons and the fermions such as the finite temperature case, one can use 
the boundary
condition associated with the $U(1)_R$ symmetry \cite{scherk}\cite{fayet}
\cite{takenaga}, in which the $U(1)_R$ charges 
are different between the bosons 
and the fermions in a supermultiplet. If we regard the factor
$\e^{-i2\pi\alpha}$ in Eq. (\ref{boundary}) as 
a boundary condition that breaks supersymmetry, our mechanism 
of supersymmetry breaking is quite similar to 
that through boundary condition associated with the $U(1)_R$ symmetry. 
If one takes this similarity seriously, one says that  
a possible physical origin of supersymmetry breaking 
through the boundary condition associated with the $U(1)_R$ symmetry
has been found. Needless to say, in order to confirm this 
statement, we need to 
clarify how the quantization ambiguity is realized in (supersymmetric) 
quantum field theory \cite{zee}.
\par
\vskip 2cm
\begin{center}
{\bf Acknowledgments}
\end{center}
\vspace{10pt}
I would like to thank Professor Jan Ambj{\rm $\phi$}rn for fruitful 
discussions, encouraging me, and critical comments on this manuscript, and 
the Niels Bohr Institute for warm hospitality. I would also like to thank 
Professor Shogo Tanimura (Kyoto University) for valuable discussions
on the quantum mechanics on $S^1$. 
\vskip 2cm

\end{document}